\pdfoutput=1
\documentclass{nature}

\usepackage{multicol}
\usepackage{multirow}
\usepackage{tabularx}
\usepackage{graphicx}
\usepackage{gensymb}
\graphicspath{{./figures/}}
\usepackage{array}
\usepackage{color}
\usepackage{amssymb}
\usepackage{upgreek}
\usepackage[font=singlespacing]{caption}
\graphicspath{{./figures/}}
\usepackage{mathptmx}
\usepackage[final]{pdfpages}

\usepackage{amsmath}
\usepackage{mathtools}

\usepackage{algorithmic}

\usepackage{array}

\usepackage{url}

\title{Direct observation of coherent energy transfer in nonlinear micro-mechanical oscillators}
\author{Changyao Chen$^{1}$, Dami\'an H. Zanette$^{2}$, David Czaplewski$^{1}$, Steven Shaw$^{3}$, and Daniel L\'opez$^{1\ast}$}

\begin{document}
\maketitle
\begin{affiliations}
\item Center for Nanoscale Materials, Argonne National Laboratory, 9700 South Cass Avenue, Argonne, IL 60439
\item Centro At\'omico Bariloche and Instituto Balseiro, Comisi\'on Nacional de Energ\'{\i}a At\'omica. Consejo Nacional de Investigaciones Cient\'{\i}ficas y T\'ecnicas. 8400 San Carlos de Bariloche, R\'{\i} o Negro, Argentina.
\item Department of Mechanical and Aerospace Engineering, Florida Institute of Technology, 150 West University Blvd., Melbourne, FL 32901

\normalsize{$^\ast$ Corresponding email: dlopez@anl.gov}
\end{affiliations}

\begin{abstract}
Energy dissipation is an unavoidable phenomenon of physical systems that are directly coupled to an external environmental bath. The ability to engineer the processes responsible for dissipation and coupling is fundamental to manipulate the state of such systems\cite{Razavy_book_2005,Braginsky_book_1985}. This is particularly important in oscillatory states whose dynamic response is used for many applications, e.g. micro and nano-mechanical resonators for sensing\cite{Chaste_nnano_2012,Decca_prl_2003} and timing\cite{time_book,Beek_jmm_2012}, qubits for quantum engineering\cite{Connell_nature_2010}, and vibrational modes for optomechanical devices\cite{Aspelmeyer_book_2014}. In situations where stable oscillations are required, the energy dissipated by the vibrational modes is usually compensated by replenishment from external energy sources. Consequently, if the external energy supply is removed, the amplitude of oscillations start to decay immediately, since there is no means to restitute the energy dissipated. Here, we demonstrate a novel strategy to maintain stable oscillations, i.e. with constant amplitude and frequency, without supplying external energy to compensate losses. The fundamental intrinsic mechanism of resonant mode coupling is used to redistribute and store mechanical energy among vibrational modes and coherently transfer it back to the principal mode when the external excitation is off. To experimentally demonstrate this phenomenon that defies physical intuition, we exploit the nonlinear dynamic response of microelectromechanical (MEMS) oscillators to couple two different vibrational modes through an internal resonance\cite{Antonio_ncomm_2012}. Since the underlying mechanism describing the fundamentals of this new phenomenon is generic and representative of a large variety of systems, the presented method provides a new dissipation engineering strategy that would enable a new generation of autonomous devices.
\end{abstract}

Micro and nano-scale mechanical resonators are examples of oscillatory systems that have been studied for decades, since they offer great flexibility to design their mechanical response and their intrinsic and extrinsic dissipation mechanisms\cite{Dykman_book_2012,Cleland_book}. For instance, expeditious energy dissipation is required in applications such as vibration isolation or switching, where mechanical motions need to be quickly damped out\cite{Okamoto_apl_2014,Ford_lambda_1999}. Conversely, small dissipation rates - or, equivalently, efficient isolation of the system of interest from its environment - are highly desired in other applications because they directly provide more stable frequency sources\cite{Beek_jmm_2012}, enhanced detection of extremely weak forces\cite{LIGO_prl_2016}, and allow for room-temperature quantum-coherent operations and sensing\cite{Wilson_prl_2009,Norte_prl_2016,Reinhardt_prx_2016}. The goal of dissipation engineering is to control the dissipative processes in which the energy stored in the system can be guided towards the natural environment or another resonant mode in a predefined manner. If the interaction with the other modes --such as an electromagnetic cavity\cite{Teufel_nature_2011,Singh_nnano_2014}, a light field\cite{Weis_science_2010}, a phononic crystal\cite{Eichenfield_nature_2009_phononic} or a mechanical vibration\cite{Mahboob_nphys_2012}-- is stronger than the interaction with the thermal bath, a new regime emerges as the system dynamics can be designed to enable distinctive behaviors. Most commonly, parametric modulation is used to activate and control the interaction, where an {\it external} modal control is applied to mediate the dynamics. Although parametric modulation can couple any two modes without any prerequisite relationship of their frequencies\cite{Teufel_nature_2011,Mahboob_nphys_2012}, hence capable of bridging distinct domains (e.g. optical modes in THz range, electrical modes in GHz range, and mechanical below MHz), the disparate frequencies typically make it difficult to achieve strong intermodal coupling. Here, we introduce a general mechanism that naturally couples mechanical modes with an intermodal coupling rate that greatly exceeds the intrinsic relaxation rate of the coupled modes. This is achieved by coupling different vibrational modes of a single oscillator through an internal resonance. Internal resonance\cite{Manevich_book_2005} (IR) capitalizes on the condition where the resonant frequencies of two distinct modes satisfy a commensurate relationship to enable strong coupling and efficient energy transfer\cite{Nayfeh_book_2008}. Operating micromechanical oscillators in the nonlinear regime, where the resonant frequency has a strong dependence on the oscillation amplitude, allows for precise tuning of the principal mode oscillation frequency just by increasing the applied driving force and thus it can be used to access the IR condition. At IR, energy initially imparted to the principal mode will be continuously exchanged between all the resonantly coupled modes and, if the energy supply is turned off, this energy will be coherently redirected toward the principal mode, effectively keeping the principal mode of the resonator oscillating. Although IR holds great promise in areas such as frequency stabilization\cite{Antonio_ncomm_2012} and energy harvesting\cite{Cottone_prl_2009,Cao_epj_2015}, it has never been utilized to engineer dissipation processes in micro and nano-resonators, partly because of the lack of definitive experimental results revealing the details of the energy transfer mechanism. 

Clamped-clamped (c-c) beam resonators are the most common and popular type used in MEMS and nanoelectromechanical (NEMS) devices: they are straightforward to fabricate at the micro and nano-scale, their fundamental dissipation processes have been thoroughly studied\cite{Timoshenko1974,Cleland_book,Ghaffari_jmems_2015}, and their nonlinear vibrational response can be precisely tailored\cite{Dou_ptrsa_2015}. As a consequence, it is possible to fabricate c-c beam resonators with a large detuning range\cite{Lifshitz_2009} that facilitates the coupling between mechanical modes. In our study, we used a single crystalline silicon c-c beam to demonstrate the IR phenomenon. The natural frequency of the principal in-plane flexural mode, $f_\text{in-plane}$, is found to be $61.4$ kHz through electrostatic actuation and detection, with a linear damping rate $\gamma_1/2\pi = 0.50$ Hz, corresponding to a quality factor ($Q$) of $122680$ (Fig. 1a). When the in-plane motion is driven with increased actuation, the large transverse amplitude leads to non-negligible elongation of the beam, causing the restoring force to vary nonlinearly with the displacement\cite{Lifshitz_2009}. This well-known Duffing nonlinearity shifts the resonant frequency upward with increasing actuation, as shown in Fig. 1b. In this case, the resonant frequency ($f_\text{res}$) corresponds to the value where the vibrational amplitude peaks as the drive frequency is swept up. For a Duffing nonlinear resonator, the increase of the resonant frequency scales quadratically with the actuation, or equivalently, with $v_\text{ac}$ in our setup (see Supplementary Information), as is shown in Fig. 1c. The resonant frequency increases quadratically with the driving strength until it reaches $f_\text{IR}=64.9$ kHz, where it saturates due to the coupling of the principal mode with a higher frequency mode through an IR. This saturation frequency matches exactly one third of a higher frequency torsional mode of the same c-c beam, whose thermomechanical noise spectrum (measured separately) is shown in Fig. 1d. The natural frequency of this torsional mode is $f_\text{torsional} = 194.6$ kHz, with a linear damping rate $\gamma_2/2\pi =2.25$ Hz (corresponding to a $Q$ of $86505$). At this tuning condition, the higher frequency (torsional) mode drains mechanical energy from the principal mode via a nonlinear inter-modal coupling mechanism: instead of building a larger flexural in-plane mode, the energy input at frequency $f_\text{IR}$ is used to excite the torsional mode at a frequency three times larger than $f_\text{IR}$ (1:3 mode coupling). 

In order to better understand the mechanism responsible for energy exchange and dynamic evolution of the IR, we put the resonator in a stationary oscillatory state through a closed-loop  configuration, and then perform time-resolved measurements of its oscillation amplitude and frequency before and after the driving force is switched off (ringdown). We apply different self-sustaining schemes that allow the in-plane motion to reach autonomous oscillations whose frequency is determined by the mechanical resonance, which can be further tuned by the feedback force through either a phase-locked loop (PLL)\cite{Antonio_ncomm_2012} or a feedback phase-delay\cite{Arroyo_epjb_2016,Chen_prl_2016}. When the resonator's frequency is outside IR (i.e., in-plane oscillation frequency different from $64.9$ kHz), the energy decay mechanism is typical of a single mode nonlinear oscillator\cite{Polunin_jmems_2016}: after the excitation is turned off, the amplitude immediately starts to decay towards zero, and the frequency decays toward the natural frequency of the principal in-plane flexural mode, $f_\text{in-plane}$ (Fig. 2a and 2b). The amplitude decay is well approximated by a simple exponential decay, with a dissipation rate of $0.68$ Hz (Fig. 2a, inset). This dissipation rate is larger than the value obtained from quasi-static frequency sweep (Fig. 1a) due to other nonlinear processes dominant at large amplitudes\cite{Polunin_jmems_2016} and frequency fluctuations\cite{Dykman_book_2012}. The temporal evolution of the instantaneous frequency during ringdown, obtained by performing fast Fourier transform (FFT) on the time-domain data, (Fig. 2b), clearly shows that the instantaneous frequency also decays from its initial value ($63.6$ kHz) towards the natural frequency of the principal in-plane flexural mode ($61.4$ kHz). The decay is also approximately exponential, with a decay rate of $1.39$ Hz, i.e. twice that of the amplitude, as expected for a Duffing nonlinearity (see Supplementary Information). When the system is brought into IR as the in-plane oscillation frequency approaches $64.9$ kHz, we find an unexpected and qualitatively different ringdown behavior: after the external excitation is removed, the amplitude and frequency of the in-plane motion remain constant for a certain period of time $t_\text{coherent}$ (Fig. 2c and 2d). During $t_\text{coherent}$, the resonator continues oscillating with a stable sinusoidal waveform as if the external energy supply were still on (Fig. 2e). In other words, it behaves as an ideal autonomous oscillator that requires neither the external sustaining feedback circuitry for power supply nor a frequency reference. Beyond $t_\text{coherent}$, the principal mode resumes the expected exponential decay, both in terms of amplitude and frequency, as in a single mode nonlinear oscillator\cite{Polunin_jmems_2016}. Figure 2f shows the dependence of $t_\text{coherent}$ on the actuation force, $v_\text{ac}$ in our setup, suggesting a practical method to control the period of time where the in-plane mode does not decay.  

A simple theoretical description of the results described in Fig. 2 can be obtained by modeling the IR condition through a 1:1 coupling of the principal nonlinear Duffing mode with a linear oscillator representing the higher frequency mode. In this case, the equations of motion for the two coupled modes are:
\begin{eqnarray}
\label{EOM}
\ddot{x}_1 + \gamma_1 \dot{x}_1 + x_1 + \beta {x^3_1} &=& J x_2 + F_\text{fb},\nonumber\\
\ddot{x}_2 + \gamma_2 \dot{x}_2 + \Omega_2^2 x_2 &=& J' x_1,
\end{eqnarray}
where $\gamma_1$, $\gamma_2$ are the dissipation rates for the respective modes, $\beta$ is the Duffing nonlinear coefficient, and $\Omega_2 \gtrsim 1$ is the normalized frequency of mode $x_2$. The modes $x_1$ and $x_2$ are linearly coupled to each other, with coupling coefficients $J$ and $J'$. Only mode $x_1$ is driven by an external feedback force $F_\text{fb}$, to compensate the energy dissipation and achieve self-sustaining oscillations. For a positive Duffing coefficient, $\beta > 0$, the oscillation frequency of mode $x_1$, $\Omega_1$, can be tuned up from its natural frequency (normalized to $1$) by increasing $F_\text{fb}$. Without inter-modal coupling ($J = J' = 0$), the coupled system reduces to two independent modes, with mode $x_2$ asymptotically at rest, and mode $x_1$ behaving as a self-sustaining nonlinear oscillator with tunable frequency\cite{Chen_prl_2016} (see Supplementary Information). For finite linear coupling, mode $x_2$ can now  be excited due to the motion of mode $x_1$. When the oscillation frequency of mode $x_1$ coincides with that of mode $x_2$, the IR condition is achieved and mode $x_2$ is then driven resonantly with strength determined by the coupling $J'$. 

The stationary solutions to equation (\ref{EOM}) determine the possible initial conditions of the ringdown evolution which, in turn, is governed by the same equations but with $F_\text{fb}=0$. For the case where both modes have small dissipation rates ($\gamma_1, \gamma_2 \ll 1$), we can apply a perturbation method to solve Eq. (\ref{EOM}), for both the steady state and the transient response during ringdown\cite{nayfeh_book_perturbation_2008,Arroyo_epjb_2016} (see Supplementary Information). Figure 3 shows the numerical solutions of equation (\ref{EOM}) for  small dissipation ($\gamma_1, \gamma_2 \sim 10^{-5}$), sufficiently large Duffing nonlinear coefficients ensuring that the principal mode amplitude is well within the nonlinear regime ($\beta \sim 10^{-3}$ to $10^{-2}$), and large coupling coefficients ($J \times J' \sim 10^{-9}$ to $10^{-7}$). The time-resolved energy decay of the principal mode is plotted in Fig. 3a, and is compared with the decay of the nonlinear resonator outside IR. The simplified model presented above qualitatively reproduces the experimental results of Fig. 2: at IR and for a finite period of time, $t_\text{coherent}$, the principal mode continues oscillating with practically constant amplitude (Fig. 3a) and frequency (Fig. 3b) after the external energy supply has been switched off. Figure 3c shows the accompanying time evolution of the higher frequency mode, indicating a rapid, non-exponential decay in its amplitude reaching toward zero at  time $t_\text{coherent}$. These results indicate a net transfer of energy from the higher frequency mode to the principal mode, since the former decays at a rate much faster than the exponential decay rate characteristic of energy dissipation toward the environmental thermal bath (dotted line in Fig. 3c). The difference between the two curves in Fig. 3c provides a direct estimation of the amount of energy transfer from mode $x_2$ to  mode $x_1$ during the period $t_\text{coherent}$. It clearly shows that a significant amount of the energy gets redirected toward mode $x_1$ instead of being dissipated to the environmental bath. For times longer than $t_\text{coherent}$, the amplitude of the higher frequency mode approaches zero, and the principal mode begins its exponential decay toward equilibrium. For the particular case of our experiment, with a 1:3 mode coupling between a flexural in-plane mode and a torsional mode, the torsional (higher frequency) mode acts as an energy reservoir for the flexural in-plane (principal) mode, by storing mechanical energy during the stationary state and transferring it back when the external feedback force is switched off. 

A closer look to the time dependence of the principal mode amplitude during $t_\text{coherent}$ shows coherent oscillations around its steady state value, with a relatively very small amplitude ($\sim 0.1 \%$) and a slowly decreasing frequency (Fig. 3d). Furthermore, the phase difference between the two modes, $\eta$, whose value is constant before ringdown, presents similar oscillations that indicates a oscillating flow of energy between the two modes (Fig. 3e). The simulated transient responses of $\eta$ shows that it oscillates around a positive value, which implies that, despite the instantaneous direction of energy flow, there is a net energy flow from mode $x_2$ to mode $x_1$. Consequently, the amplitude of mode $x_2$ decays rather abruptly, while the amplitude of mode $x_1$, remains - on the average - practically constant. This dynamical exchange breaks down as soon as mode $x_2$ exhausts its energy, and from then on, mode $x_1$ behaves as a single mode nonlinear oscillator. 

%The origin of such fine structure can be explained as follows: when the external force is turned off, the forces that maintained the stationary oscillation become unbalanced. This unbalance, however, is rather small since the oscillator has large quality factor, and the self-sustaining force necessary to maintain the oscillations is very small as compared with other involved forces involved (i.e., linear and cubic elastic forces: see Supplementary Information). At the same time, the external action that maintained the synchronization between the two modes disappear, and the two modes will behave as two coupled, weakly damped oscillator with different natural frequencies and no mechanism able to sustain their synchrony and to counteract the energy dissipation. However, the main mode (mode $x_1$) has not left far from the previous force-balanced state, therefore it will take some time for it to move away. 

While our experimental setup cannot directly measure the coherent amplitude oscillations shown in Fig. 3d due to the limitation of the measurement bandwidth, we do see their presence in the time-resolved frequency spectrum before and during $t_\text{coherent}$. Figure 4 shows the temporal evolution of the resonator's frequency as it transits from steady-state to ringdown in a device that is driven to a strong IR condition with larger feedback force $F_\text{fb}$. During steady-state we find two sidebands flanked around the main peak in the power spectrum (regime I), that during $t_\text{coherent}$ (regime II), both sidebands evolve towards the frequency of the principal in-plane flexural mode, merging with it at $t=t_\text{coherent}$. When the frequency of mode $x_1$ is exponentially decaying (regime III), there is no evidence of sidebands because there is no coupling between the vibrational modes and thus no exchange of energy. The emergence of these sidebands is a direct consequence of the energy change (cf. Fig. 3e), whose value quantifies the exchange rate $\gamma_{ex}$. For the particular case presented in Fig. 4a, we obtained a $\gamma_{ex} \sim 800$ Hz, which is almost three orders of magnitude larger than the intrinsic damping rate of each of the coupled modes ($\sim 1$ Hz). By performing similar frequency measurements at different driving forces, a linear correlation between $t_\text{coherent}$ and $\gamma_{ex}$ is obtained (Fig. 4a, right inset). The data indicates that a $\gamma_{ex} > 200$ Hz is needed in order to have a finite $t_\text{coherent}$ and the larger the $\gamma_{ex}$, the larger the $t_\text{coherent}$.

The experimental and numerical results discussed above demonstrate that mode coupling can be used to engineer the intrinsic relaxation phenomena of nonlinear oscillators. In particular, when different vibrational modes are coupled through an IR, the exchange of energy between modes could happen orders of magnitude faster than the exchange of energy with the external environmental bath (Fig. 4b). Under these conditions, nonlinear resonators can sustain, for a finite period of time, stable oscillations without external energy supply. The dissipation engineering concept presented in this work could be applied to a wide range of MEMS and NEMS oscillators whose performance is limited by the electrical noise in the feedback circuit\cite{Feng_nnano_2008}. MEMS and NEMS resonators oscillating without external power should be ideal devices to identify the ultimate stability limit imposed by thermomechanical noise\cite{Sansa_nnano_2016}. The possibility to control the energy exchange rate between coupled modes creates a testbed to validate theories of thermalization of nonlinear systems out of equilibrium\cite{Dykman_jetp_1975,Midtvedt_prl_2014} and resonant nonlinear friction\cite{Shoshani_unpublished}. In atomically thin NEMS resonators, where the nonlinear dynamic response can be easily achieved\cite{Chen_pieee_2013}, mode coupling can have a significant effect on the relaxation process toward equilibrium\cite{Bachtold_unpublished}.

It is worthwhile emphasizing the manifold role of nonlinearity in the occurrence of the phenomena studied here. In the first place, nonlinearity is responsible for the coupling between different oscillation modes, enabling the exchange of energy that determines their mutual influence. At the same time, the upshift of the oscillation frequency as the driving force increases in amplitude makes it possible that the nonlinear oscillator reaches the condition of IR - an effect which is absent in harmonic oscillators. Finally, the presence of an interval of coherence just after the driving force is switched off has also to be ascribed to nonlinearity. The combination of beating and exponential decay that characterizes the dynamics of coupled \emph{linear} oscillators is in fact not able to account for such temporary stability of the oscillation amplitude in the absence of an external action.

\newpage
\noindent \textbf{References}
\vspace{2 mm}
\bibliography{reflib1}

\newpage

\begin{figure}
\caption{
\textbf{Internal resonance (IR) in a nonlinear MEMS oscillator.} 
\textbf{a}. Linear resonance of the in-plane oscillation mode measured with open-loop setup, DC bias V$_\text{DC} = 7$ V, and alternating current (ac) voltage actuation $v_\text{ac}= 40$ $\mu$V. Inset: finite element simulation of the in-plane mode shape (CoventorWare). The length and thickness of the c-c beam are 500 $\mu$m and 10 $\mu$m, respectively. The central portion consists of 3 parallel beams, each with width of 3 $\mu$m.
\textbf{b}. The vibrational amplitude of the in-plane motion as the external excitation frequency is swept up (open-loop), with excitation amplitude $v_\text{ac}$ increasing from $1$ to $19$ mV at $2$ mV steps. At large $v_\text{ac}$, the  vibrational amplitude spectrum shows an asymmetrical line-shape that skews towards  high frequencies, due to the positive Duffing nonlinearity. As $v_\text{ac}$ increases, the frequency at which the vibrational amplitude peaks, $f_\text{res}$, also increases, until reaching $64.9$ kHz. 
\textbf{c}. Measured $f_\text{res}$ with different $v_\text{ac}$. For $v_\text{ac} \lesssim 7.5$ mV, $f_\text{res}$ follows a quadratic dependence on $v_\text{ac}$. For $v_\text{ac} \gtrsim 7.5$ mV, $f_\text{res}$ stabilizes at $64.9$ kHz, which equals to one third of $f_\text{torsional}$. Notice that all the voltages are rms values.
\textbf{d}. Thermomechanical noise spectrum of the torsional mode measured with optical interferometry (see Supplementary Information for details). 
}
\end{figure}

\begin{figure}
\caption{
\textbf{Ringdown responses outside and inside IR.}
\textbf{a}. Oscillation displacement of the in-plane mode before and after the external drive is turned off (at time equals $0$ sec), when its oscillation frequency is outside the IR condition. The inset shows the extracted envelope of the displacement, plotted in logarithmic scale. The black line is a fit with an exponential decay, with a decay rate of 0.68 Hz. The external excitation $v_\text{ac}$ is 6.3 mV.
\textbf{b}. Temporal frequency response of the oscillation during the same period of time. The power spectrum at each nominal time $t_i$ is obtained by performing a noverlapping fast Fourier transform of the time domain data in a narrow window of 16 ms around $t_i$. The inset shows the extracted instantaneous frequency offset from the principle in-plane frequency, with fitting an to exponential decay (black line). The fitted decay rate is 1.39 Hz.
\textbf{c}. Oscillation displacement of the in-plane mode before and after the external drive is turned off, when its oscillation frequency is inside the IR condition. For the first 108 ms after the ringdown starts, the envelope of the oscillation remains practically constant. The inset shows the extracted oscillation envelop during ringdown, with an exponential decay fit for the $t > t_\text{coherent}$ portion. 
\textbf{d}. Temporal frequency response of the oscillation during the same period of time, showing constant frequency oscillations during $t_\text{coherent}$. The inset shows the extracted instantaneous frequency offset from the principle in-plane frequency, with fitting to an exponential decay (black line) for the $t > t_\text{coherent}$ portion. The fitted decay rate is 1.28 Hz. For $t < 0$, the self-sustaining motion is driven by an external PLL that outputs a single-frequency signal. For $t \in (0,t_\text{coherent})$ the transient response is only controlled by the system dynamics and two sidebands flanking the main frequency peak appear (see main text and Supplementary Information for details).
\textbf{e}.  Zoomed-in view of the oscillation of ringdown of \textbf{c}, indicating clean and stable sinusoidal oscillations (the time is offset to 50 ms). 
\textbf{f}. $t_\text{coherent}$ obtained with different steady-state drive $v_\text{ac}$. The red line is a theoretical fit with the model described in Equation (1) (see Supplementary Information for details).
}
\end{figure}

\begin{figure}
\caption{
\textbf{Modeling the ringdown response of a nonlinear oscillator.} 
\textbf{a}. Simulated transient responses of the vibrational amplitude decay (i.e. envelope of the oscillation displacement) of a single mode nonlinear oscillator (blue) and a nonlinear oscillator at internal resonance (red). Note that for the nonlinear oscillator at IR, the vibrational amplitude does not decay during $t_\text{coherent}$ after the external excitation is removed at $t = 0$. 
\textbf{b}. Simulated transient responses of the instantaneous frequency offset, $\Omega_1 - 1$, for the two cases presented in \textbf{a}. After the external excitation is removed, the instantaneous frequency for the single mode nonlinear oscillator (blue) decays exponentially towards zero, while for the nonlinear oscillator at IR (red), the instantaneous frequency stays constant.
\textbf{c}. Temporal evolution during ringdown of the higher frequency mode $x_2$ at IR: the amplitude (red line) decays much faster than the expected exponential decay (dashed black line), indicating faster energy transfer to mode $x_1$ than to the environmental bath. The shaded area indicates the amount of energy transferred from mode $x_2$ to mode $x_1$.
\textbf{d}. Time dependence of the principal mode ($x_1$) amplitude during the period $t_\text{coherent}$, revealing small oscillations ($\sim$ 0.1 \%) around its initial value, with a decreasing frequency. 
\textbf{e}. Simulated phase difference $\eta$ between the two coupled modes during the same period of time as in \textbf{d}. It shows the presence of coherent oscillations around a positive average value of $\eta$, indicating a net energy transfer from from mode $x_2$ to mode $x_1$. 
}
\end{figure}

\begin{figure}
\caption{
\textbf{Time scale of the energy exchange dynamic at IR.}
\textbf{a}. Temporal evolution of the instantaneous frequency before (regime I) and during (regime II and III) ringdown. For $t < 0$, the self-sustaining motion is driven by a linearly feedback signal with appropriate phase-delay, instead of using a PLL setup as the one used to obtain the data presented in Figure 2. In regime I, there are two sidebands flanking the main steady-state frequency (64.9 kHz, IR) that indicate the energy exchange rate between the main and higher frequency modes (a single power spectrum is shown in left inset). During  $t_\text{coherent}$ (regime II), the main frequency stays constant and the sidebands merge toward the main frequency peak, and finally disappear at the end of regime II. Regime III represents the normal decay for a single mode nonlinear oscillator and thus shows no evidence of sidebands. Right inset: extracted $\gamma_\text{ex}$ plotted against corresponding $t_\text{coherent}$, showing a linear correlation between them. 
\textbf{b}. Schematic representation of the energy flow during $t_\text{coherent}$ (regime II): there is a net energy flow from the high frequency torsional mode to the in-plane principal mode at a rate $\gamma_\text{ex}$ $\gg$ $\gamma_1, \gamma_2$. This large difference in  relaxation rates causes the in-plane principal mode to maintain stable oscillations even after the external power supply has been switched off.
}
\end{figure}

\begin{addendum}
\item The authors like to thank Mark Dykman, Jeffrey Guest, Aftab Ahmed, Ori Shoshani, Pavel Polunin, Thomas Kenny, Adrian Bachtold, Andreas Isacsson, and Vladimir Aksyuk for critical discussions.  Use of the Center for Nanoscale Materials at the Argonne National Laboratory was supported by the U.S. Department of Energy, Office of Science, Office of Basic Energy Sciences, under Contract No. DE-AC02-06CH11357. S.S. is supported by NSF grant 1561829 and funds from Florida Institute of Technology.
\item[Competing Interests] The authors declare that they have no competing financial interests.

\item[Author contributions] C.C., D.C. and D.L. conceived and designed the experiments, C.C. and D.C. performed the experiments and analyzed the data, D.H.Z., S.S. and C.C. performed the modeling and numerical simulations, C.C. and D.L. co-wrote the paper. All authors discussed the results and commented on the manuscript.
 \item[Correspondence] Correspondence and requests for materials
should be addressed to \\
D.L.~(email: dlopez@anl.gov).

% \item[Additional information] Supplementary information accompanies this paper at\\ www.nature.com. Reprints and permission information is available online at\\ http://npg.nature.com/reprintsandpermissions/.
% Correspondence and requests for materials should be addressed to D.L..
\end{addendum}

\includepdf[pages=-]{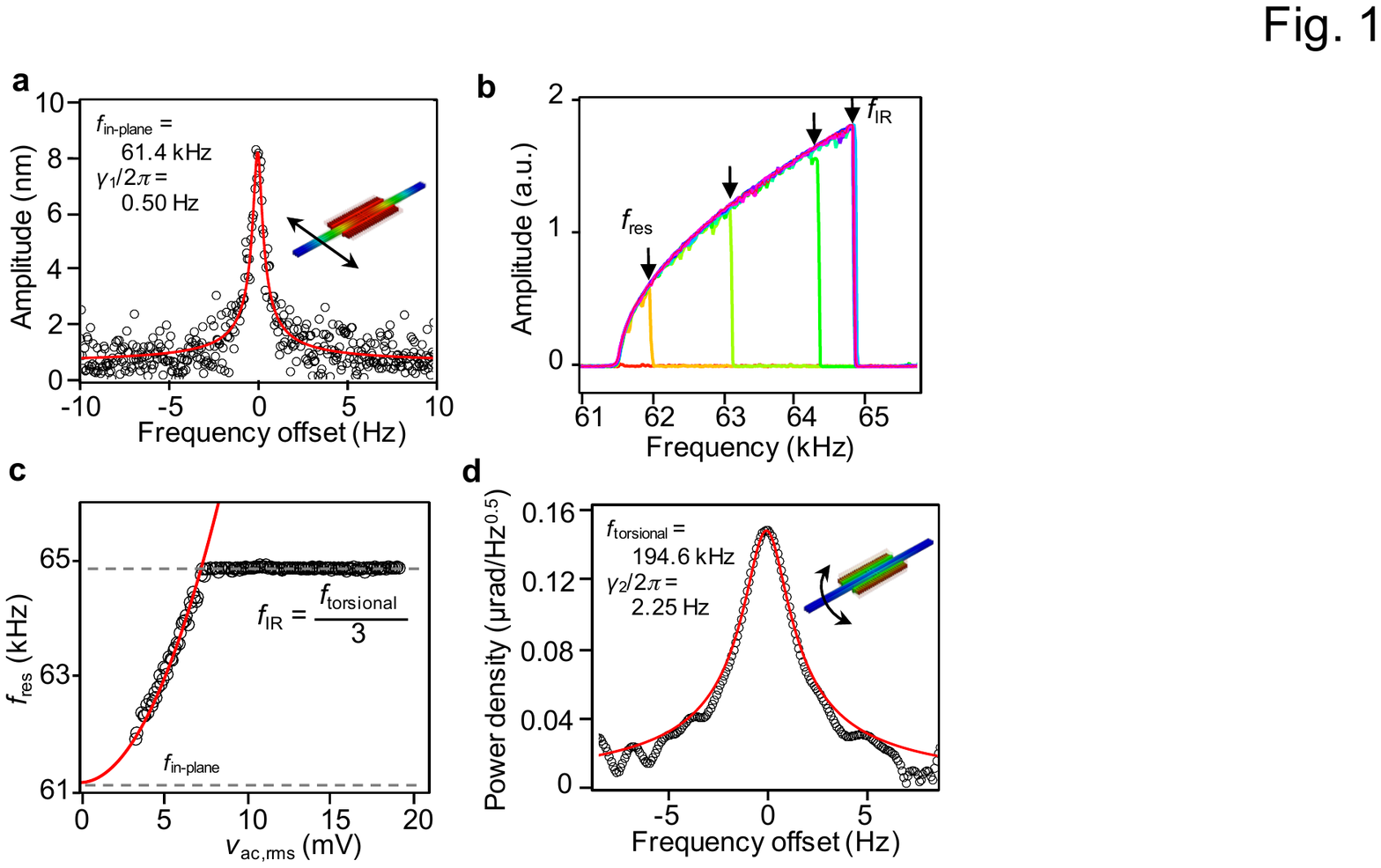}

\end{document}